\documentclass[runningheads]{llncs}

\usepackage[T1]{fontenc}
\usepackage{booktabs}
\usepackage{graphicx}
\usepackage{amssymb}
\usepackage[sectionbib,square,numbers]{natbib}
\usepackage{etoolbox}
\usepackage[hidelinks]{hyperref}
\usepackage{amsmath}
\makeatletter
\patchcmd{\@spthm}{\phantomsection}{}{}{}
\makeatother

\usepackage[nameinlink]{cleveref}

\begin{document}

\title{Introducing ORKG ASK: an AI-driven Scholarly Literature Search and Exploration System Taking a Neuro-Symbolic Approach}
\titlerunning{ASK: an AI-driven Scholarly Literature Search and Exploration System}

\author{Allard Oelen\inst{1}\orcidID{0000-0001-9924-9153} \and
Mohamad Yaser Jaradeh\inst{2}\orcidID{0000-0001-8777-2780} \and
S\"oren Auer\inst{1,2}\orcidID{0000-0002-0698-2864}}

\authorrunning{Oelen et al.}

\institute{TIB – Leibniz Information Centre for Science and Technology, Hannover, Germany 
\email{\{allard.oelen,auer\}@tib.eu}\and
L3S Research Center, Leibniz University of Hannover, Hannover, Germany
\email{jaradeh@l3s.de}}

\maketitle

\begin{abstract}
As the volume of published scholarly literature continues to grow, finding relevant literature becomes increasingly difficult. With the rise of generative Artificial Intelligence (AI), and particularly Large Language Models (LLMs), new possibilities emerge to find and explore literature. We introduce ASK (Assistant for Scientific Knowledge), an AI-driven scholarly literature search and exploration system that follows a neuro-symbolic approach. ASK aims to provide active support to researchers in finding relevant scholarly literature by leveraging vector search, LLMs, and knowledge graphs. The system allows users to input research questions in natural language and retrieve relevant articles. ASK automatically extracts key information and generates answers to research questions using a Retrieval-Augmented Generation (RAG) approach. We present an evaluation of ASK, assessing the system's usability and usefulness. Findings indicate that the system is user-friendly and users are generally satisfied while using the system. 


\end{abstract}

\section{Introduction}
\label{sec:introduction}
Analyzing scholarly literature is a key aspect of research. 
However, due to the ever-increasing body of scholarly publications, finding scholarly literature becomes increasingly difficult~\cite{landhuis2016scientific}. 
Consequently, finding literature consumes a substantial portion of researchers' time~\cite{niu2010national}. 
Because of the recent developments in generative Artificial Intelligence (AI), and specifically Large Language Models (LLMs), new possibilities arise to extract knowledge from scholarly articles, helping researchers to find relevant literature in the flood of publications.

\begin{figure}[t]
\centering
    \includegraphics[width=\textwidth]{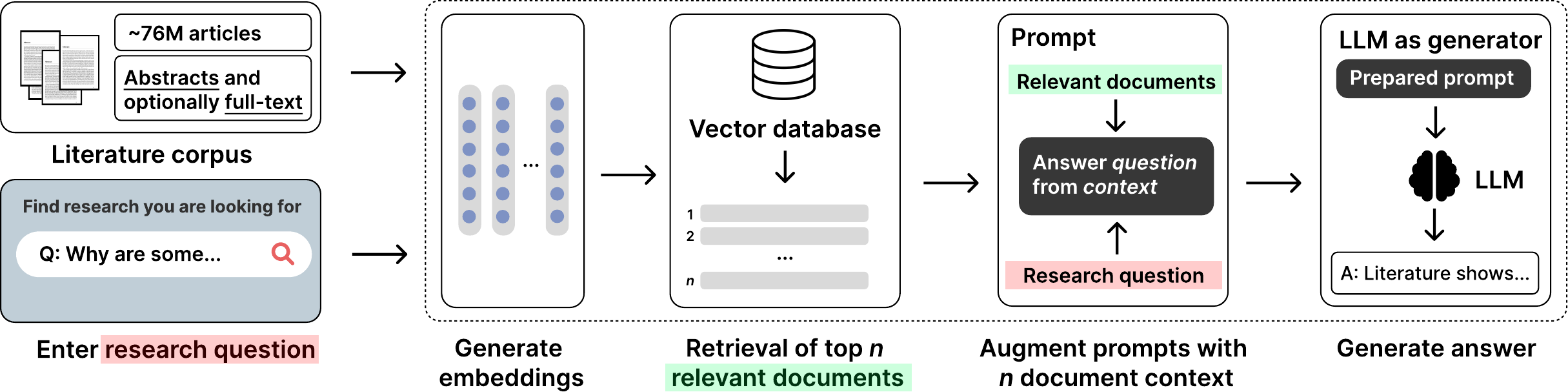}
    \caption{Explainer depicting our RAG (Retrieval-Augmented Generation) approach for scholarly search. 
    The \textit{Retrieval} step ranks articles by their relevance to the question. The \textit{Augmented} step injects the previously retrieved context in the prompt. The \textit{Generation} step prompts the LLM and displays the answer.}
    \label{fig:teaser}
\end{figure}

In this article, we present ORKG ASK (Assistant for Scientific Knowledge), hereafter referred to as ASK, a next-generation scholarly search and exploration system. ASK aims to provide support to researchers in finding relevant scholarly literature.
ASK takes a Neuro-Symbolic approach which consists of three key components, namely Vector Search and LLMs for the neural aspect and Knowledge Graphs (KGs) for the symbolic part. We build upon our previously presented work where we demonstrated the basic ASK infrastructure~\cite{oelen2024orkg}. In this paper, we expand on our previous work by providing an in-depth explanation of the approach, technical details of the implementation, and a extensive evaluation. In brief, ASK functions as follows: a user of ASK formulates their information needs as a research question. Afterward, a list of relevant articles is displayed. For each article, an automatically extracted answer for the previously asked question is displayed to the user. Finally, the symbolic aspect ensures users are able to narrow down the search space by providing semantic filters. This provides both the precision of symbolic approach and the flexibility of a neural approach.
ASK is running as a publicly available production service online.\footnote{\url{https://ask.orkg.org}} 

\begin{figure*}[tb]
    \centering
    \includegraphics[width=\textwidth]{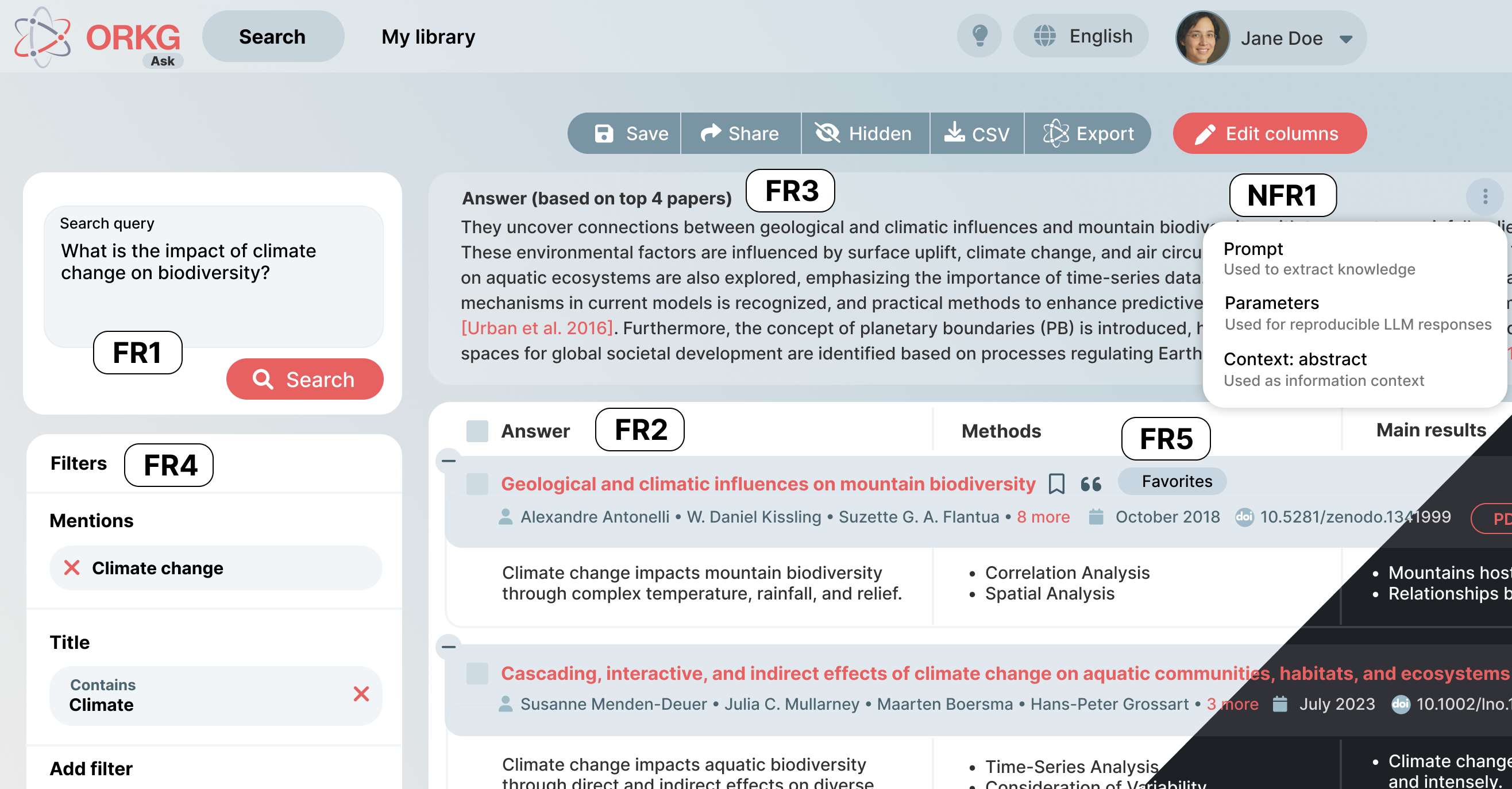} 
    \caption{Screenshot of the ASK search results page. The nodes with (N)FR correspond to the implementation of the (Non-)functional requirements, as listed in \autoref{table:requirements}.}
    \label{figure:screenshot}
\end{figure*}

The system takes a Retrieval-Augmented Generation (RAG) approach to support the previously described workflow. RAG~\cite{rag2020} is commonly used to intertwine LLM extractions with information retrieval systems, as depicted in~\autoref{fig:teaser}. Firstly, the Vector Search component ranks documents based on their relevance (Retrieval) for a research question. Secondly, relevant context is collected (i.e., the paper abstract and, if available, full-text) from the previous step (Augmented). Finally, the LLM generates answers and displays this to the user (Generation). A screenshot showing the ranked articles, search query, and generated LLM responses is displayed in~\autoref{figure:screenshot}. This work introduces the following contributions: i) presents an LLM-supported open-source scholarly literature search and exploration service, ii) describes a scholarly RAG system leveraging LLMs, vector search, and KGs, and iii) provides insights from the design and development process, supported with an evaluation.

\section{Related Work}
\label{sec:related-work}
The landscape of scholarly search systems can be categorized into two groups, domain-agnostic and domain-specific systems. Prominent examples of domain-agnostic systems include Google Scholar, Semantic Scholar, and Scopus~\cite{gusenbauerWhichAcademicSearch2020}.
Well-known domain-specific systems include PubMed 
for the medical domain and ACM Digital Library  
for the computer science domain.
At a high level, these systems function similarly: relevant articles are returned for a set of user-provided keywords. 
A new generation of scholarly search systems tries to automatically extract relevant information from articles. 
Among others, those systems include Elicit, Consensus, and Scispace~\cite{bolanosArtificialIntelligenceLiterature2024}. Those systems are similar in that sense that they are not open source, making it difficult to determine what models are employed and what underlying data corpus is used. This makes the reproducibility of results, for example for a systematic literature review, a challenging task. 
Additionally, the trustworthiness of generated responses by Large Language Models (LLMs) becomes increasingly problematic when the underlying technologies are not transparent. To the best of our knowledge, the previously mentioned systems typically use a RAG (Retrieval-Augmented Generation)~\cite{rag2020} approach, as also mentioned by Bolanos et al.~\cite{bolanosArtificialIntelligenceLiterature2024}. This roughly resembles the approach of ASK. However, a notable difference between ASK and the other systems is the open nature of ASK. All source code is openly available online, and we clearly communicate which data corpus we use and which models we employ. This makes ASK more suitable for reproducible literature searches. 

LLMs gained a lot of traction after the seminal work of Vaswani et al. which paved the way for transformer models like BERT, RoBERTa, and more scientifically-oriented models like SciBERT~\cite{beltagy2019scibertpretrainedlanguagemodel}.
Models then quickly started growing in size and capabilities such as GPT3~\cite{brown2020languagemodelsfewshotlearners}.
From this point on, very large models started to show impressive performance across a vast spectrum of language tasks. 
Despite their impressive capabilities, LLMs face several key challenges. 
There are concerns about the use of LLMs in fields requiring high precision, such as healthcare~\cite{shen2023chatgpt} and law, where inaccuracies can have serious consequences. 
A particular issue is the phenomenon of ``hallucinations'', where models generate statements that, while plausible, are entirely fabricated and lack factual basis~\cite{bang2023multitask}.

In the context of the scholarly domain, LLMs need to be more accurate for widespread usage~\cite{susnjak2024automatingresearch}.
CORE-GPT~\cite{pride2023coregpt} is an effort to combine LLMs like ChatGPT with open-access research to provide a more trustworthy and credible scientific question answering.
Furthermore, Van Dis et al. highlighted that researchers need to pay extra attention when using LLMs for research purposes specifically when applying them to literature comprehension and summarization tasks~\cite{van2023chatgpt}. Our work with ASK positions itself in the middle, trying to transparently show where the answers came from and at the same highlight to users that the answers are automatically generated via a language model and as such need to be reviewed by a human. ASK bridges the missing part of search systems which is the natural language expression and connects it with the advanced capabilities of LLMs to get the best of both worlds. 

\section{Approach}
\label{sec:approach}
In this section we present our approach for scholarly search and discuss the system requirements, which essentially cultivated in the ASK system.

\subsection{RAG for Scholarly Search} 
\label{ssec:rag}

For a scholarly search system, parametric knowledge of LLMs should be limited and not used as a main source of information.
Parametric knowledge is the knowledge that models encode within their vast number of parameters, and with it, LLMs are able to answer questions.
Relying solely on such knowledge can lead to hallucinations and the generation of inaccurate information.
For the aforementioned reasons, ASK relies on Retrieval-Augmented Generation (RAG)~\cite{rag2020} to combine the parametric knowledge of the models and the non-parametric knowledge stored in vector stores to generate accurate and related text.

\subsubsection{Non-Parametric Memory}
Also referred to as Semantic Search.
Non-parametric memory is a part of the procedure that extends the knowledge reservoir of pre-trained language models to the requirements of individual applications. This type of memory provides various benefits to scholarly search:
i) Customizability: the knowledge base contains only the items of interest and as such direct the LLM to answer only in relation to documents indexed within the vector store.
ii) Updatable knowledge base: since there is no need to retrain the language model with every new document added, new documents can be easily indexed and added to the already-existing knowledge base.
iii) Complex filtering capabilities: the vector store also offers the flexibility to further filter or refine search results based on metadata or other available criteria.

In order for the retriever component to work, first a set of documents needs to be processed and indexed inside a vector store. The vector store is populated with a semantical representation of documents, via embeddings. ASK uses the Nomic embedding model\footnote{In particular ``nomic-embed-text-v1.5'' which utilizes Matryoshka Representation Learning~\cite{kusupati2024matryoshkarepresentationlearning}.} which has an embedding size of 768 and a context window sequence length of 8K token.
The choice of the embedding model is based on its advanced multilingual capabilities, efficient parameter utilization via MoEs, and its long context handling ability.
Finally, the collection of documents retrieved is then passed down to the parametric memory component for further processing.

\subsubsection{Parametric Memory}
The parametric memory component solely relies on the language model itself. 
Using both parametric and non-parametric memories side-by-side has multiple benefits for LLM-based applications:
i) Tailored responses: rather than posing a general query to the model and expecting an answer, the model now receives the query and the context in which it is supposed to look up the answer.
ii) Reduced hallucinations: LLMs are notorious for hallucinating content~\cite{perkovic2024hallucinations}. With specifying the context, the model is forced to rely on the text that exists within its prompt and not within its own parametric knowledge.
iii) Instruction following: the parametric knowledge within the models allows for custom instructions to generate apt responses depending on the use case.

ASK utilizes custom-made prompts containing placeholders that take pieces of information from the non-parametric memory and are then used by the LLM. 
ASK uses the small variant of Mistral LLM\footnote{ASK uses Mistral 7B Instruct v0.2 with no sliding-window attention.}~\cite{jiang2023mistral7b}. The usage of a relatively small model reduces the required computational resources and loading times, while still being well capable of following instructions. The Mistral model has a 32K tokens context window which is suitable for passing the full text of articles and getting specific answers. Inferencing with LLMs can be resource-intensive and is usually the bottleneck when it comes to performance in production systems.
For this particular reason, a caching mechanism is employed with the parametric memory. Caching is applied on partial hits (i.e., for single cells), making it possible to return single responses partially from the cache and partially from the LLM. The implemented caching mechanism reduces the loading times significantly and prevents calling the LLM, which in turn benefits computational efficiency. 

\subsection{System Requirements}
To provide guidance during development, we formulate a set of system requirements, as listed in~\autoref{table:requirements}. The requirements are divided into functional and non-functional requirements. For brevity reasons, we list the high-level requirements only. The functional requirements focus on literature search, information extraction, and the ability to filter and organize information. The two most important non-functional requirements are the reproducibility options and the focus on barrier-free access via various accessibility features.

\begin{table*}[t]
\centering
\caption{List of functional and non-functional system requirements, outlining the high-level key concepts to guide the system development.}
\label{table:requirements}
\resizebox{\textwidth}{!}{%
\begin{tabular}{@{}l|p{2.5cm}|p{6cm}|p{6cm}@{}}
\toprule
\textbf{ID} & \textbf{Title} & \textbf{Requirement} & \textbf{Rationale} \\
\midrule
\multicolumn{4}{l}{\textit{Functional requirements}} \\
\midrule
\textbf{FR1} & Literature search & The system shall allow users to find scholarly literature for research questions. & To provide a scholarly search and exploration system. \\
\textbf{FR2} & Information extraction & The system shall display automatically extracted information from found literature. & To ensure users get a quick overview of the literature so relevancy can be assessed. \\
\textbf{FR3} & Answer synthesis & The system shall provide a summarized answer to research questions. & To provide a clear answer to the research question based on a set of articles. \\
\textbf{FR4} & Result filtering & The system shall allow users to set filters for finding related semantic concepts. & To narrow down the search space and provide a more fine-grained search. \\
\textbf{FR5} & Bibliography manager & The system shall provide a bibliography manager to store related literature. & To ensure collections can be stored and to allow importing existing articles. \\
\midrule
\multicolumn{4}{l}{\textit{Non-functional requirements}} \\
\midrule
\textbf{NFR1} & Reproducibility & The system shall always produce reproducible responses. & To ensure the system is suitable for scholarly research and is transparent to users. \\
\textbf{NFR2} & Accessibility & The system shall follow accessibility guidelines to ensure accessibility for all users. & To ensure that users with disabilities can use all features. \\
\textbf{NFR3} & Usability & The system shall be easy-to-use and can be operated with a minimal learning curve. & To provide an alternative to existing scholarly search systems. \\
\textbf{NFR4} & Maintainability & The system shall follow established coding standards to facilitate maintainability. & To ensure the system can be employed as a sustainable service. \\
\textbf{NFR5} & Interoperability & The system shall be interoperable with existing bibliography managers.  
& To ensure literature can be imported and exported to existing systems.  \\
\bottomrule
\end{tabular}%
}
\end{table*}

\section{Implementation}
\label{sec:implementation}
In this section, we present the implementation details to realize the ASK system.
We present the functional and non-functional requirements, the LLM setup for various use cases, the dataset used to populate the index of the search component, and technical details about the implementation.

\subsection{Requirements Realization}
In~\autoref{figure:screenshot}, a screenshot depicts the ASK interface. We will now discuss how the previously listed system requirements are implemented within this interface. 

\subsubsection{Functional Requirements}
The literature search (FR1) is implemented by providing a large search box on the homepage from where users can get started by entering a research question. 
The research result page shows the question and a list of results ranked by relevance (\autoref{figure:screenshot} node FR1).
The LLM-extracted answer is displayed in node FR2, alongside additional columns that are extracted as well. Users can modify those columns to extract specific information by clicking the \textit{Edit columns} button. For the answer synthesis, a summarized answer is displayed at node FR3. Citations within this summarized answer point to the results listed below. Results can be filtered in the box displayed at node FR4. Finally, items can be added to a bibliography collection by clicking the bookmark icon as displayed in node FR5.

\begin{figure*}[t]
    \centering
    \includegraphics[width=\textwidth]{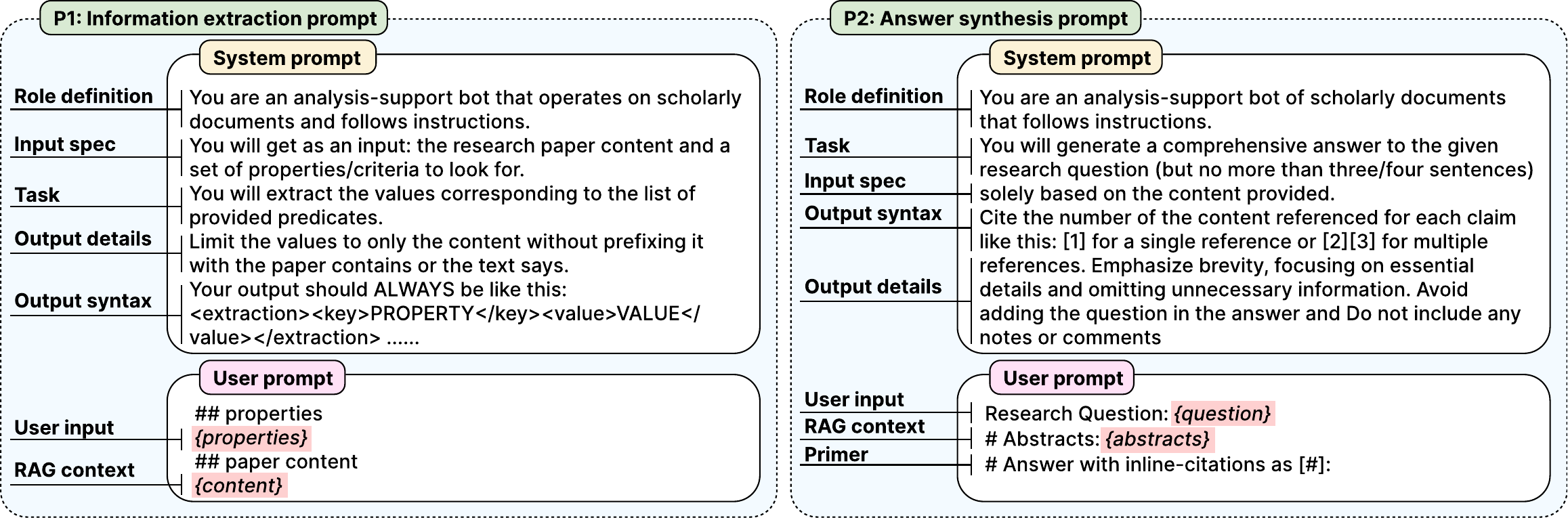}
    \caption{Sample prompts for different RAG use cases within ASK. The system prompts provide the instructions to the LLM (Prompt P1 trimmed for brevity reasons). The user prompt includes the user input and RAG context. Values highlighted in red are placeholders used to inject user values into the prompt. Additionally, P2 uses a primer to improve the answer of the LLM.}
    \label{fig:prompts}
\end{figure*}

\subsubsection{Non-Functional Requirements}
\label{sec:accessibility}
In addition to being open source, ASK also provides data to ensure results are reproducible (NFR1). We created the reproducibility menu (see \autoref{figure:screenshot}, node NFR1) that provides for all LLM-generated content: i) the prompts, i) the model, iii) the parameters (such as temperature, seed, etc.), and iv) the context used for the generation (full text or abstract). The transparency helps users to better assess the results' correctness and enables them to reproduce the same results themselves. Among other features, this sets ASK apart from other services, which are often proprietary and lack transparency. 
Accessibility is another key aspect of ASK. Any user, including those with disabilities, should be able to use the system barrier-free (NFR2). We integrated various features to facilitate accessibility. 
Firstly, the interface is responsive, making sure that the interface is usable at large zoom levels, benefiting users with visual impairments. As an additional benefit, the service can also be operated on different screen sizes, such as tablets and mobile phones. A dark mode (as displayed on the right bottom of~\autoref{figure:screenshot}) replaces all light colors with dark ones, and can be enabled to reduce eye strain. 
Secondly, ARIA attributes are added to facilitate screen reader usage, benefiting users with visual impairments~\cite{craig2009accessible}. Finally, the interface is internationalized, meaning that the service can be operated in different languages and different regions.
The LLM responses are provided in multiple languages as well, opening up new methods to find related literature in the preferred language of the user, and making literature search more inclusive.
In the end, the various accessibility features benefit all users.

The interface is designed to look intuitive and modern benefiting usability (NFR3). 
To ensure the system is maintainable (NFR4) and can be operated as a production service, it is implemented using the latest technologies and code standards. Details about the implementation are described in \autoref{sec:technical-details}. To make the system interoperable with other reference managers (NFR5), we adopted the Citation Style Language (CSL)\footnote{\url{https://citationstyles.org}} throughout the system.

\subsection{LLM Setup}
\label{ssec:llm-setup}

A construct of an LLM chain is implemented.
A chain is the combination of three components: i) Prompt, ii) Model, and a iii) Parser.
The prompt is an aggregation of the system and user prompt for a particular task (see \autoref{fig:prompts}).
Before the invocation of the model, the relevant information retrieved by the non-parametric memory is injected and formatted into the prompt.
Secondly, the model is the LLM and potentially any LoRAs~\cite{hu2021loralowrankadaptationlarge} that need to be applied to the language model\footnote{ASK does not implement any LoRAs at the moment. However, this technique can be integrated to further customize the model for domain-specific use cases.}.
Lastly, a parser is a function that gets called on the output (i.e., the response of the LLM) and is then parsed, sanitized, and formatted to be used in other parts of the application. 
We note that we did not need to employ custom-trained or fine-tuned LLMs for ASK at this stage. As the LLM is mostly used as a means to perform information extraction, and in turn text generation, a fine-tuned model would not necessarily result in higher-quality results. 

\subsection{Dataset}
ASK uses the CORE~\cite{Knoth2023-zi} dataset of open-access research papers as the basis of its indexed corpus. 
The CORE data is automatically crawled from open-access repositories and publisher websites.
This means that there are quality-related issues that require some curation before any ingestion operation takes place.
Before indexing the CORE data in the vector store of ASK, a pre-processing phase was implemented to choose, based on a set of heuristics, which items and articles are suitable to be added.
This process involves checking if the articles have valid titles and abstracts (i.e., non-empty and have a length greater than a threshold).
The abstracts proved to be the most impacting factor within this process.
The data import process is a continuous process as the CORE data is growing with time and other sources are also integrated within the ASK system. In total, we imported 76.4M articles from the CORE dataset, excluding items that do not follow the previously mentioned requirements. Of the imported articles, 36.9\% have a DOI and 25\% have full-text available. 

In addition to the CORE data, we imported a subset of BMBF-conform (German Federal Ministry of Education and Research) research reports related to autonomous driving, containing approximately 310 reports.\footnote{\url{https://ask.orkg.org/search?query=&filter=AND[0][source][inList][0]=TIB\%2520Forschungsberichte\%2520Autonomes\%2520Fahren}} The ASK service is operated by the German National Library of Science and Technology (TIB) and by importing this dataset, we demonstrate how ASK can be leveraged to explore the library's special collections. In the future, we plan to import more of such special collections. 

\subsection{Technical Details}
The system is developed using a microservices setup\footnote {Server configuration: 1TB of RAM, 15TBs of SSD storage, 128 CPU cores, and seven GPU cards (Nvidia L4 4x24GB, Nvidia L40S 2x46GB, and Nvidia H100 1x80GB).} and is available as a free online service. The services is divided into the frontend and backend. 
\label{sec:technical-details}
The frontend is written in TypeScript with React. It uses the Next.js framework, adopting the server components paradigm where suitable. Furthermore, it uses Tailwind for styling and HeroUI as a component library. The use of standardized technologies increases the maintainability, as described in NFR3. The frontend is available as open-source software and is published with a permissive MIT license.\footnote{\url{https://gitlab.com/TIBHannover/orkg/orkg-ask/frontend}} 

The backend is mainly written in Python leveraging the FastAPI framework. The backend adopts a modular approach where each functionality is in its own module, which improves maintainability and extensibility, as described in NFR4. Other components to serve the language models, vectorized documents, and cache items are part of the backend but are written in different languages and are used as turn-key solutions. The source code of the backend and various components are available publicly under the MIT license.\footnote{\url{https://gitlab.com/TIBHannover/orkg/orkg-ask/backend}} Furthermore, ASK utilizes Qdrant\footnote{\url{https://qdrant.tech}} as a vector store and the TGI\footnote{\url{https://huggingface.co/text-generation-inference}} engine for serving LLMs and inferencing in production. The containers are managed via Podman. 

\section{Evaluation}
We now discuss the system evaluation, which is divided into two parts: subjective user evaluations and objective data analysis. 

\label{sec:evaluation}
\subsection{Subjective Evaluation}
ASK is publicly released as a scholarly information retrieval service and is being actively used by researchers. To gather feedback from real-world system users, we integrated a lightweight feedback collection component into the user interface. 
The feedback component appears on question pages.
The component consists of two different sets of questions. 
The first set evaluates the helpfulness, correctness, and completeness of the displayed question and its answers. The second set of questions asks users about their general feedback on the ASK system. The questions consist of two standardized and unmodified UMUX-Lite~\cite{lewis2013umux} questions and one to assess whether users are satisfied with ASK. User satisfaction is another commonly used method to assess usability. The operational feedback is collected on a running basis, previous results consisting of a small number of responses (approximately 3\%) have been published already in a demo article~\cite{oelen2024orkg}. 
Participation in the operation feedback questionnaire is on an opt-in basis and users can close the feedback popup if they do not wish to participate. 

\begin{figure}[t]
\centering
    \includegraphics[width=0.74\textwidth]{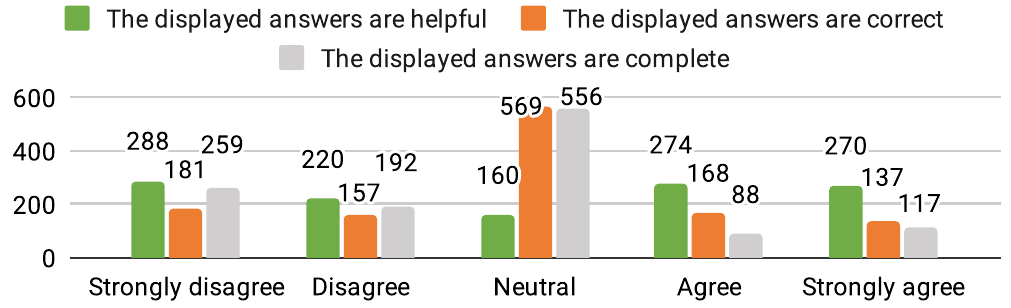} 
    \caption{Question specific results for operational feedback collection. }
    \label{figure:question-specific-results}
\end{figure}

The question-specific form has been filled out 1,212 times in the period from June 15, 2024, until January 15, 2025. Based on browser fingerprinting, it was completed by 1,032 different users. The results are displayed in \autoref{figure:question-specific-results}.
As can be observed, the results of the helpfulness of answers vary among users. This means that participants experience different levels of relevance for the answers, which can be explained by their expectations of the system. The correctness of answers is voted as more neutral. Meaning that users might had difficulty assessing the correctness, or thought answers were neither fully correct nor incorrect. Finally, the results for completeness are similar to correctness, indicating that most users had no strong opinion about the completeness of the answers.

\begin{figure}[t]
\centering
  \begin{minipage}[c]{0.62\textwidth}
    \vspace{0pt}
    \centering
    \includegraphics[width=0.70\textwidth]{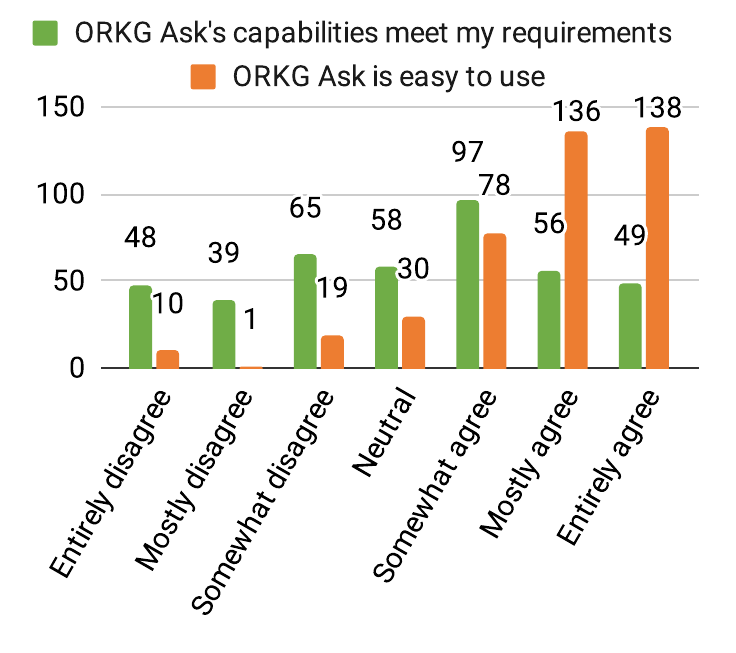} 
    \caption{UMUX-Lite results with a score of 65.7.}
    \label{figure:umux-results}
  \end{minipage}
  \begin{minipage}[c]{0.35\textwidth}
  \vspace{0pt}
    \centering
    \includegraphics[width=0.8\textwidth]{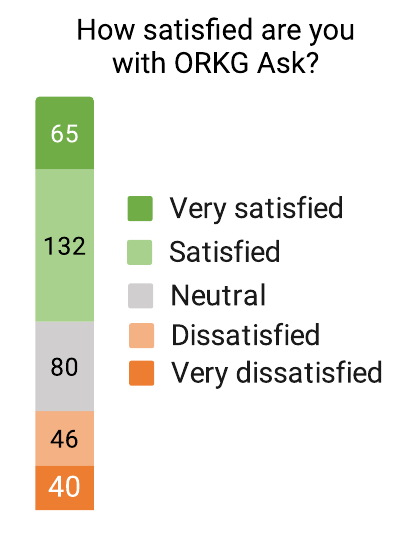} 
    \caption{General user satisfaction of ASK. }
    \label{figure:user-satisfaction}
\end{minipage}
 \end{figure}

The results of UMUX-lite questions to assess the general usability of the system are displayed in \autoref{figure:umux-results}. A total of approximately 443 users filled out this evaluation. As the questions were optional, the number of users differs slightly from the numbers in the figure. The numbers result in a calculated UMUX score of 65.7 on a scale of 0-100, where higher scores indicate better usability. 
Incomplete partial responses were discarded in the final UMUX calculation, resulting in 409 included responses. As the results show, ASK does not always meet the users' requirements. 
However, the majority of users do agree that the system itself is easy to use. This indicates the design decisions to make the system easy to use have proven to be effective. Finally, \autoref{figure:user-satisfaction} displays the user satisfaction outcomes. In total, 363 users answered this question. As can be observed, average user satisfaction leans toward more positive than negative opinions. 

\subsubsection{User Experiment}
In addition to the operational feedback, we conducted a small-scale user experiment to compare ASK to an established literature search system, specifically Google Scholar. We were particularly interested in the perceived task load differences between ASK and Google Scholar. For this, we designed a within-subject study where a total of 9 participants had to answer a set of four predefined research questions, two per condition. Most of the participants are engaged in academic research, have either a master or PhD degree, and have used ASK before, but were not involved in the development of ASK. The majority of participants (7 out of 9) searches for academic articles at least on a weekly basis. To counteract sequence bias, the two conditions were evaluated by participants in random order. To answer the questions, the participants had to use at least two references per answer, the references had to be provided by the search system. Additionally, they had to manually verify the correctness of the answer from the source article, consequently, they were not allowed to copy-paste LLM-generated answers. For each condition, they had to indicate their perceived task load, measured with an unweighted NASA Task Load Index (TLX) scale~\cite{Hart2006a}. Additionally,  the time to answer a question was recorded. 

The results of the comparisons between the ASK and Scholar search condition are displayed in \autoref{table:experimental-results}. As can be observed, the ASK condition has a considerably lower perceived task load compared to the Scholar condition. Regarding the required time, there was an outlier that took more than 113 minutes to answer the questions via ASK, while only taking 24 minutes to answer via Scholar. For completeness, the timing results are displayed with and without the outlier. When the outlier is disregarded, the required time to answer questions with ASK is lower than with Scholar. Although ASK and Google Scholar might not be directly comparable, making the conclusions less definitive, the comparison gives insights into the future direction of providing alternatives to established scholarly search systems.

\begin{table}[b]
\centering
\caption{Showing a comparison between the ASK and Scholar condition regarding task load and required time. Task load is listed as percentage and time in seconds. }
\label{table:experimental-results}
\resizebox{0.55\textwidth}{!}{%
\begin{tabular}{@{}l|rr|rr@{}}
\toprule
 & \textbf{ASK} &  & \textbf{Scholar} &  \\ \midrule
 & \multicolumn{1}{r|}{\textit{Mean}} & \textit{SD} & \multicolumn{1}{r|}{\textit{Mean}} & \textit{SD} \\
Task load (TLX) (\%) & \multicolumn{1}{r|}{26.76} & 16.65 & \multicolumn{1}{r|}{61.3} & 16.59 \\
Time (s) & \multicolumn{1}{r|}{984.41} & 456.67 & \multicolumn{1}{r|}{1241.39} & 496.36 \\
Time with outlier (s) & \multicolumn{1}{r|}{1628.79} & 1979.78 & \multicolumn{1}{r|}{1267.49} & 470.86 \\ \bottomrule
\end{tabular}%
}
\end{table}

\subsection{Objective Evaluation}

\begin{table}[t]
        \caption{Web analytics data and user interaction statistics of the ASK production service. Measured from May 15, 2024, until February 1, 2025. }
\label{table:user-interaction-statistics}
    \centering
    \resizebox{0.87\textwidth}{!}{%
    \begin{minipage}[t]{0.3\textwidth}
        \centering
        \begin{tabular}{@{}l|r@{}}
\toprule
\textbf{Analytics} &  \\ \midrule
Visits & 74,145 \\
Returning visits & 26,354 \\
Pageviews & 219,189 \\
Duration visit & 4:01m \\
Bounce rate & 3\% \\
 &  \\
 \bottomrule
        \end{tabular}
    \end{minipage}
    
    \begin{minipage}[t]{0.36\textwidth}
        \centering
\begin{tabular}{@{}l|r@{}}
\toprule
\textbf{Events} &  \\ \midrule
Queries asked & 67,949 \\
Downloads & 7,595 \\
Outlinks & 19,723 \\
Custom filters added & 723 \\
Custom columns added & 415 \\
Load more (1 page) & 5,067 \\
Load more (2 pages) & 2,149 \\
Load more (\textgreater 2 pages) & 5,010 \\ \bottomrule
\end{tabular}
    \end{minipage}
      
    \begin{minipage}[t]{0.3\textwidth}
        \centering
        \begin{tabular}{@{}l|r@{}}
\toprule
\textbf{Device usage} &  \\ \midrule
Desktop & 76,9\% \\
Smartphone & 21,2\% \\
Other & 1,9\% \\
 &  \\ 
  \bottomrule
        \end{tabular}
    \end{minipage}
    
    }
\end{table}
To gain insights into how to system is used, we collected data using web analytics. We analyzed user interaction data over the period starting from May 15, 2024, until February 1, 2025. Analytics data was recorded via Matomo\footnote{\url{https://matomo.org}} and used browser fingerprinting to distinguish between users. Additionally, specific events in the interface were logged to determine how frequently they were used. A summarized overview of the analytics and user interaction statistics is presented in \autoref{table:user-interaction-statistics}. Visits are defined as users visiting the service who have not visited a page in the last 30 minutes. The bounce rate of 3\% is rather low, indicating that users are actually using the system, and not just visiting a single page and then leaving the service.
A considerable amount of users are using ASK on devices other than desktops. This indicates that the service indeed works well for different screen sizes, part of our accessibility requirement NFR2. 

The recorded events gain more insights into what features are actually being used. The downloads and out-links relate to the number of articles that are being visited from ASK. This includes following links to the PDF, publisher landing pages, or data repositories. The number of added custom filters is low. 
Also, the number of columns added is rather low, meaning that most users did not use the functionalities to extract custom data from articles.
The low number of custom extractions either means that users were already satisfied with the default extracted properties, or they did not understand how to use the feature. Finally, the load more numbers show the number of times a user clicks the ``Load more'' button at the bottom of the page, indicating they are interested in finding more related work.

\section{Discussion}
\label{sec:discussion}
ASK is meant as a scholarly search system, helping researchers to find and explore scholarly literature. Although ASK indeed also answers research questions while performing a search, first and foremost ASK is a scholarly information retrieval system. Therefore, finding literature is the main objective of ASK. We consider the question-answering feature as a means to find related work, not as an objective by itself. This means that the previously discussed RAG method is a key aspect of our approach and using LLMs in isolation, even when pre-trained or fine-tuned on scholarly articles, would not be sufficient to meet our search goal. Therefore, in the user experiment, we specifically focused on comparing ASK to a scholarly search system and not a question-answering system. 

As previously discussed, when employing LLMs, one has to keep in mind that LLMs tend to hallucinate. Hallucinations cannot be completely eliminated, even when using models with a higher number of parameters (i.e., more powerful and capable models). Indeed, for the use case of ASK, hallucinations can also occur, but to a large extent do not pose major challenges. As the question-answering capability is secondary to the main goal of information retrieval, users do not solely rely on the extracted information to find relevant literature. As mentioned, it is only a means to assess the literature's relevance. After potentially relevant articles are discovered, users are expected to perform rigorous analysis of the listed work, as is also expected when using traditional scholarly search systems. To communicate this to users, a warning message is displayed in the interface, informing users that all information needs to be manually checked. 
	  
The evaluation presents the results of operational feedback and data collected from a production environment. Consequently, data is collected in an uncontrolled environment, meaning that individual responses and data might be inaccurate or incomplete. For example, the users' intentions of filling out the evaluation forms are unknown, which could be only for testing purposes or without thoroughly reading the questions. Also, for the feedback, no demographics were collected, limiting the possibility of an in-depth analysis of the results. Browser fingerprinting was used to distinguish among different users, which means that unique users cannot be fully accurately determined. Therefore, it should be noted that the evaluation results might contain data from the same users, even though they are reported as different users. However, we do consider the results to be relevant nevertheless, and when aggregated, provide valuable insights. 

Finally, we will discuss future work. The usefulness of the approach as a search system heavily depends on the quality and the size of the underlying literature corpus. We plan to extend our literature repository by including more articles from different sources, and by parsing content semantically, for example by performing author name disambiguation, all contributing toward more semantic search. As mentioned in~\autoref{sec:related-work}, ASK sets itself apart from similar existing services by providing an open-source service and focusing on literature search reproducibility. A quantitative performance comparison with these services is out of scope for this work, but is an interesting future research direction. ASK's transparency features provide a key advantage over these services, but a comparison regarding the quality of the responses provides helpful insights into other aspects of our approach. 

\section{Conclusion}
\label{sec:conclusion}
In this work, we presented ASK, a scholarly search and exploration system. We examined how AI can be leveraged for literature searches by exploring the direction of using vector search and LLMs to provide active support to users while conducting a literature search. 
ASK showcases how such a literature search system can operate. Furthermore, we focused on the ability of LLMs to provide value to researchers while performing a literature search. To this end, we leveraged a RAG (Retrieval-Augmented Generation) approach to find and explore scholarly literature. 
Using a RAG approach, LLMs are used for information extraction and text generation, making a large literature corpus explorable using AI technologies. The RAG approach provides a literature search where source information is more easily traceable while partially mitigating common LLM limitations, such as hallucinations and limited context sizes. Finally, we investigated whether AI-driven tools, such as ASK, are potential alternatives to already established scholarly search tools. 
By presenting the ASK service and its respective user study, we captured researchers' attitudes toward the new approach and concluded there is indeed potential and interested audience for such new tools. 

\subsubsection*{Acknowledgements}
\begin{sloppypar}
This work was co-funded by NFDI4DataScience (ID: 460234259), and by the TIB Leibniz Information Centre for Science and Technology. We want to thank the entire ORKG team for their contributions to the ORKG platform, including research and development efforts.
\end{sloppypar}

\bibliographystyle{splncs04}
\bibliography{refs-manual}

\end{document}